# A high-speed and high-resolution 3D velocity map imaging system with a neuromorphic camera

Sulaiman Abubakar, Blessed Oguh, Akaninyene Udoikono, Hackim Musah, Suk Kyoung Lee and Wen Li*

Department of Chemistry, Wayne State University, Detroit, MI

wli@chem.wayne.edu

## Abstract

An improved camera-based 3D-velocity map imaging (VMI) system was demonstrated with a neuromorphic camera. We show a machine vision event-driven camera can achieve an unprecedented latency of 2 microseconds when detecting single ion/electron with a microchannel plate imaging detector, which boosts the speed of the momentum imaging system to more than 100 kHz, while the system with a conventional frame camera is limited to a few kHz. The high spatial resolution of the camera offers additional performance boost while maintaining excellent multi-hit capability and a high time resolution.

## Introduction

The camera-based three-dimensional (3D) velocity map imaging (VMI) detection system[1-7] employs a fast frame camera to detect the position of an ion/electron hit on a microchannel plate (MCP)/phosphor screen imaging detector, just like in a conventional 2D VMI system[8]. Additionally, to achieve 3D detection, the time-of-flight (TOF) of the same ions/electron hit is characterized using a timing device such as a high-speed digitizer to sample the time-resolved electric signals from the MCP pick-off that arises from the hits. The system offers a high-performance alternative to conventional time- and position-sensitive detectors such as delay-line detectors[9], cross-wire detectors[10] and wedge-strip detectors[11, 12]. The system has been successfully applied in various studies such as coincidence electron-ion imaging[13-15], attosecond spectroscopy[16-18] and photoemission and photo-desorption from surfaces[19-21].

While the 3D VMI system is perfectly suited to work with laser systems running at a repetition rate of a few kHz, limited by the frame rate of high-speed frame cameras, a need for higher speed has recently risen due to the fact that high repetition rate lasers running at 100 kHz have become quite accessible thanks to the advancement in high power fiber laser and pulse compression technology. A high-speed camera capable of this frame rate does exist but the short acquisition time due to the large data rate, a limited resolution and a low sensitivity render it ineffective in the 3D VMI system. A frame-based camera outputs a complete frame containing all pixels, regardless of their values, even if most pixels are zero and carry no useful information. However, an event-driven camera has a fundamentally different imaging mode which only measures the change of the light intensity on a pixel and then outputs a stream of pixels that experience such a change with a timestamp (x, y, t). In this way, the data rate is significantly reduced. Depending on the resolution/latency of the pixel timestamp, the effective frame rate can be very high (a few hundred kHz or higher). We note the latency in event driven cameras is defined as the time lag between the actual event time and the event timestamp. Recently, a 3D VMI system employing the TPX3cam demonstrated an event rate approaching 1 Mhits/s due to the superior 1.56 ns timestamp resolution[22]. However, TPX3cam has a limited spatial resolution (256×256 pixels), which limits the final momentum resolution when employed in a 3D VMI setup. A high-resolution version is currently under development which uses four timepix3 chips and can reach 512×512 pixels. Another potential high-resolution alternative is a machine vision neuromorphic camera, which has already reached a resolution of 1280×720 pixels in a Sony/Prophesee IMX636 sensor. The timestamp of such a camera has a resolution of 1 μs. However, it has been shown that such a resolution does not translate to 1 MHz effective frame rate. The reason lies in the different timestamping methods: in TPX3Cam, the timestamping takes place on the pixel, while in machine vision neuromorphic camera, the events are processed off-pixel in the peripheral logic circuits. This leads to a significant latency, which can amount to a few hundred μs. A more problematic issue is that such a latency does not have a fixed value: it varies significantly depending on the amount of light and the electronic band-width of the pixel. Therefore, the jitter of the latency can be as large as the latency. This renders the 1 μs timestamp resolution irrelevant[23]. A recent work used a machine-vision neuromorphic camera in a 2D VMI setup and reported a measured pixel latency of 200-300 μs ( iniVation, DAVIS346), which corresponded to an effective frame rate of 2 kHz[24]. The gain in terms of frame rate was minimum over existing frame cameras.

In this work, we show by tuning the pixel performance parameters (biases) and picking the earliest timestamp within a cluster of activated pixels representing a single ion/electron hit, an unprecedented 2 μs latency can be achieved with a low-cost neuromorphic camera. This boosts the speed of the camera-based 3D VMI system to more than 100 kHz while simultaneously increasing its spatial resolution. We also show the detection sensitivity of a commercial event camera can match that of a low-noise CMOS cameras and thus a machine vision event camera can be an ideal replacement for a frame camera in the camera-based 3D VMI system.

**Experimental Setups**

Two apparatus were used in this study. One is a gas-phase 3D VMI setup[13] and it has three main parts: the laser system, the ultrahigh vacuum (UHV) chamber, and the 3D VMI detection system. Briefly, laser pulses of ~30 fs were generated by a two-stage Ti:sapphire amplified laser system (KM Labs, Red Dragon, 800 nm, 1 mJ per pulse). The pulses were further compressed in a hollow core fiber compressor. The final pulse duration was measured using the dispersion scan (D-scan) technique[25, 26] and found to be about 5 fs. The compressed pulses were then focused into a UHV chamber operating at a base pressure of about $10^{-9}$ torr, where strong field photoionization of the target molecules took place. A cold molecular beam was formed by supersonic expansion of the sample gas into the vacuum chamber, passing through a skimmer placed between the source chamber and the main interaction chamber. At the interaction region, the molecular beam crossed the focused few-cycle pulses, producing photoelectrons and photoions. These charged particles were accelerated toward their respective detectors by the inhomogeneous electric field of the VMI electrostatic lens. Each detector was made of a pair of microchannel plates (MCPs) coupled to a phosphor screen of P47. The electric signal arising from ion/electron hit on the detector were picked off by an AC coupling circuit and digitized using National Instrument PXIE 5114 (for ions) and PXIE5183 high speed digitizers. In the original 3D-VMI, the light from the phosphor screen was recorded using a Basler CMOS frame camera (Aca720-520um), which has a spatial resolution of 260 by 260 pixels at 1000 frames per second. In the present work, we added an event camera based on IMX 636 (Sony/Prophesee) event sensor. The camera was manufactured by IDS (UE-39B0XCP-E). For benchmark purpose, we placed the two cameras side by side behind the phosphor screen of the ion detector (Fig. 1a). The IDS camera has a native resolution of 1280 by 720 pixels, more than ten time higher than that of the frame camera. For both cases, the cameras are triggered using a delay generator (Stanford Research Systems DG535), which in turn was triggered by a fast photodiode monitoring the laser pulse. Note in an event camera, the trigger signal is only for synchronization purpose, it does not start/stop frame exposures as in a frame camera.

The Ti:Sapphire laser runs at 1 kHz and cannot reach a higher repetition rate. To test the ability of the event camera and the detection system at high speed, a second VMI machine/laser was employed. This setup was designed for studying photoemission from surfaces and was equipped with a picosecond laser (Photonic Industries, RX3-MWB) that can run at 100 kHz[19, 21, 27]. The laser beam (355nm, 1 μJ) was focused onto a thin film of perovskites nanoplatelets, and the produced electrons were accelerated and momentum focused onto an imaging detector (Photonis, APD). The same event camera was used to register the hit positions on the phosphor screen while a digitizer (National Instrument, PXIe5160) was used to measure the time-of-flight of electrons.

**Hit Detection Algorithm**

Because the output of an event camera is different from that of frame camera, a new algorithm/program is needed to process the camera data to identify electron/ion hits, each of which may include tens to hundreds of activated pixels. We developed a new program to decode the raw byte stream from the camera, group pixel events into ion hits, and associate each hit with a laser shot via a hardware trigger signal. The camera outputs a continuous binary stream in the Prophesee EVT3 format. EVT3 is Prophesee's 16-bit, vectorized event-data encoding format, designed to efficiently represent pixel events while reducing data redundancy. With Prophesee provided library, the data were decoded to extract ON pixel events (increasing brightness) and laser trigger timestamps, which share the same hardware clock. Isolated background noise events are removed by an activity noise filter that retains only pixels with at least one spatially adjacent neighbor firing within a coincidence window $\tau$. Surviving pixels are grouped into connected components by a member-anchored breadth-first search algorithm with the same time window. This window is chosen to capture the full pixel arrival spread of a single ion flash, which includes dim periphery pixels that arrive up to ~100 μs after the bright core due to brightness-dependent threshold crossing time (time-walk) in IMX636 sensor, while preventing merging of consecutive shots at 100 kHz. Each connected component is then subdivided into individual hits by burst segmentation: a new hit is declared when a dense burst of ≥10 pixel events is detected within 3 μs, the local event rate jumps by a factor of ≥2 relative to the preceding window, and the burst onset falls within the laser gate. This resolves same-spot pile-up at 100 kHz. Once the cluster associated with each ion/electron hit was identified, the hit position was calculated as the center-of-mass of all pixels. As for the timestamp, we adopted the earliest timestamp in the cluster as the timestamp for the ion/electron event. Finally, each accepted hit is assigned to a laser shot by matching its earliest timestamp to the nearest preceding trigger timestamp. This can now be used to synchronize the hit position with the high resolution TOF captured by the digitizer to yield the 3D momentum information of each hit. Recorded quantities per hit are centroid position (x, y), cluster size ($n_{pixels}$), first pixel timestamps ($t_{first}$), latency ($t_{first}-t_{trigger}$), etc. An optional per-pixel output records the individual coordinates, timestamps, and trigger-relative times of every pixel in each hit, enabling per-pixel time-walk analysis. The program was written using C++ and has achieved real-time acquisition even at 100 kHz repetition rate thanks to the reduced data throughput of the event camera.

## Results and Discussion

Fig. 1b shows the distribution of raw pixel latency associated with detected ion events. The latency is calculated as the difference between the pixel timestamp and the closest preceding trigger timestamp. The distribution is not symmetric: it rises quickly but decay slowly. This is consistent with the fact that the latency is strongly dependent on the brightness of the flash. What is surprising is that the latency distribution is quite narrow at ~10 μs level, which is much better than the manufacturer specification (100 μs@1000 lux). This can be explained by the following consideration: the light illumination level of 1000 lux is from constant environmental illumination. A 20% change of brightness (a typical contrast change value to activate an event pixel) at this level require a significant time duration to accumulate and this becomes the largest contribution to the specified latency. However, in an imaging detector employed in VMI, the light flashes from the phosphor screen arrives in a very short time span (<100 ns for P47 phosphor). A brightness change of 20% from almost zero light background can take place very quickly and thus suppresses the latency significantly. Therefore, the use case in VMI strongly favors a reduced latency as long as the pixel electronics have high enough bandwidth to respond to the fast change. For the IMX636 sensor, this turns out to be the case. The sensor offers tuning of the speed using the low pass filter (bias_fo), which controls the bandwidth cutoff for high frequencies. In this study, we used a value of 40 to achieve a good speed and thus a low latency while keeping the noise low. We note the event pixel in IMX636 measures the brightness on logarithm scale.

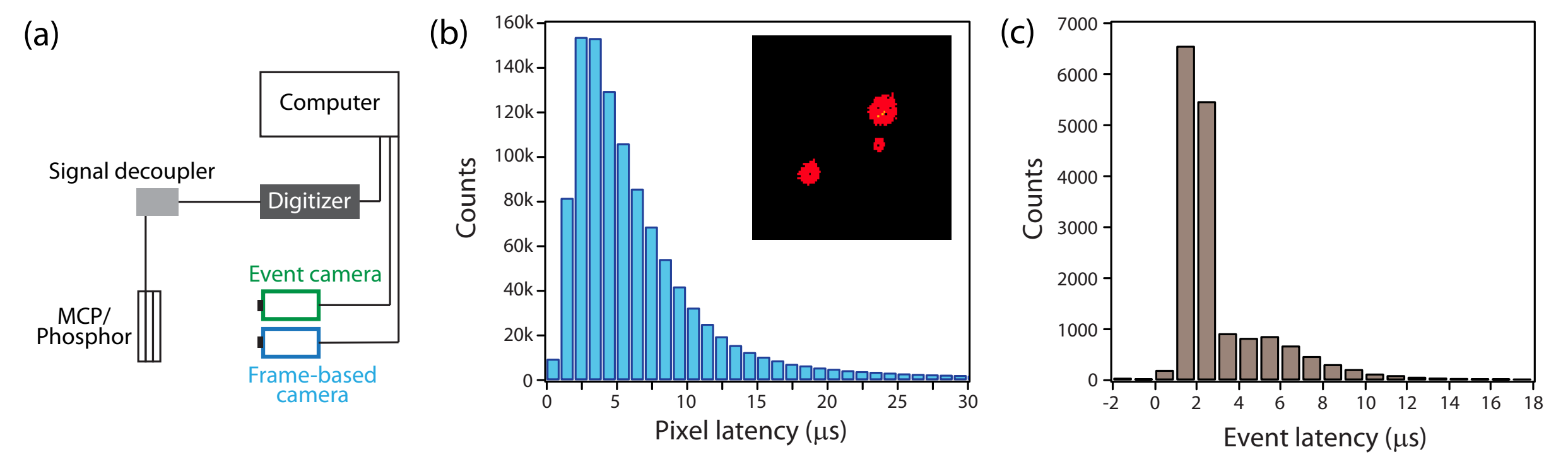

Figure 1: A scheme of experimental setup and latency characterization (a) a simplified test setup for a 3D VMI detection system. (b) A distribution of latency of all activated pixels. The inset shows clusters of activated pixels of three ion events. (c) A distribution of latency of all events, obtained by picking the earliest pixel latency among the pixels within one event.

Further improvement on latency is achieved by choosing the earliest timestamp within the cluster that is associated with one electron/ion event. Fig. 1(c) shows the results. The distribution has significantly narrowed with the main peak being only two bins (2 μs) wide and with minor contribution up to 8 μs. This is an outstanding result and suggests a potential boost of speed of the system to more than 100 kHz. We note this latency distribution is from detecting a single mass ion ($H_2^+$) on the ion detector. To further verify the latency performance, we detected multiple mass ($H_2^+$, $Ar^+$, $CO_2^+$ etc) with the event camera and the digitizer. Here the latency distribution becomes a TOF spectrum. We can now compare the two spectra (Fig. 2a). Although at a lower resolution, the event camera TOF clearly resembles the true TOF. This shows that the achieved low latency allows the machine vision event camera to become a low-resolution mass spectrometry camera, which is remarkable and might be useful by itself.

Two important requirements on camera performance needs to be considered. The first one is the sensitivity and the second is the capability to measure the brightness of event. For a camera to be suitable in the camera-based 3D VMI, its sensitivity must be high enough to detect the dim phosphor flashes, at least matching that of detecting electric signal from MCP by a digitizer, in which the signal is often amplified ten to hundreds of times by a preamplifier. This has put a stringent benchmark on the sensitivity of a camera and renders most machine vision cameras ineffective due to their high readout noise. To assess the detection efficiency of the event camera, we put it in a side-by-side comparison with the Basler frame camera, which is known to have excellent readout noise (2-3 electrons) and has been widely employed in the camera-based 3D

VMI by the Li group. Fig. 2b shows the comparison for 1000 laser pulses in sequence. Note that all three devices (Basler frame camera, IDS event camera and digitizer) are synchronized to the same laser pulses and detect the same events. Indeed, the result show the event camera has a very similar detection efficiency as the highly sensitive frame camera and the MCP pick-off scheme. A 1-2% difference in overall detection efficiency was observed between the frame and the event camera, qualifying the event camera.

The second requirement on the capability of measuring brightness arises from the fact that in camera-based 3D VMI, multi-hit events from the camera and the digitizer are correlated by the brightness of the camera events and the peak heights in the digitizer traces. While brightness (pixel value) comes natively in frame camera, it is not a direct output of event cameras. Each pixel output (x, y, t, and p, p being the polarity, ON or OFF) does not contain any information on the brightness.

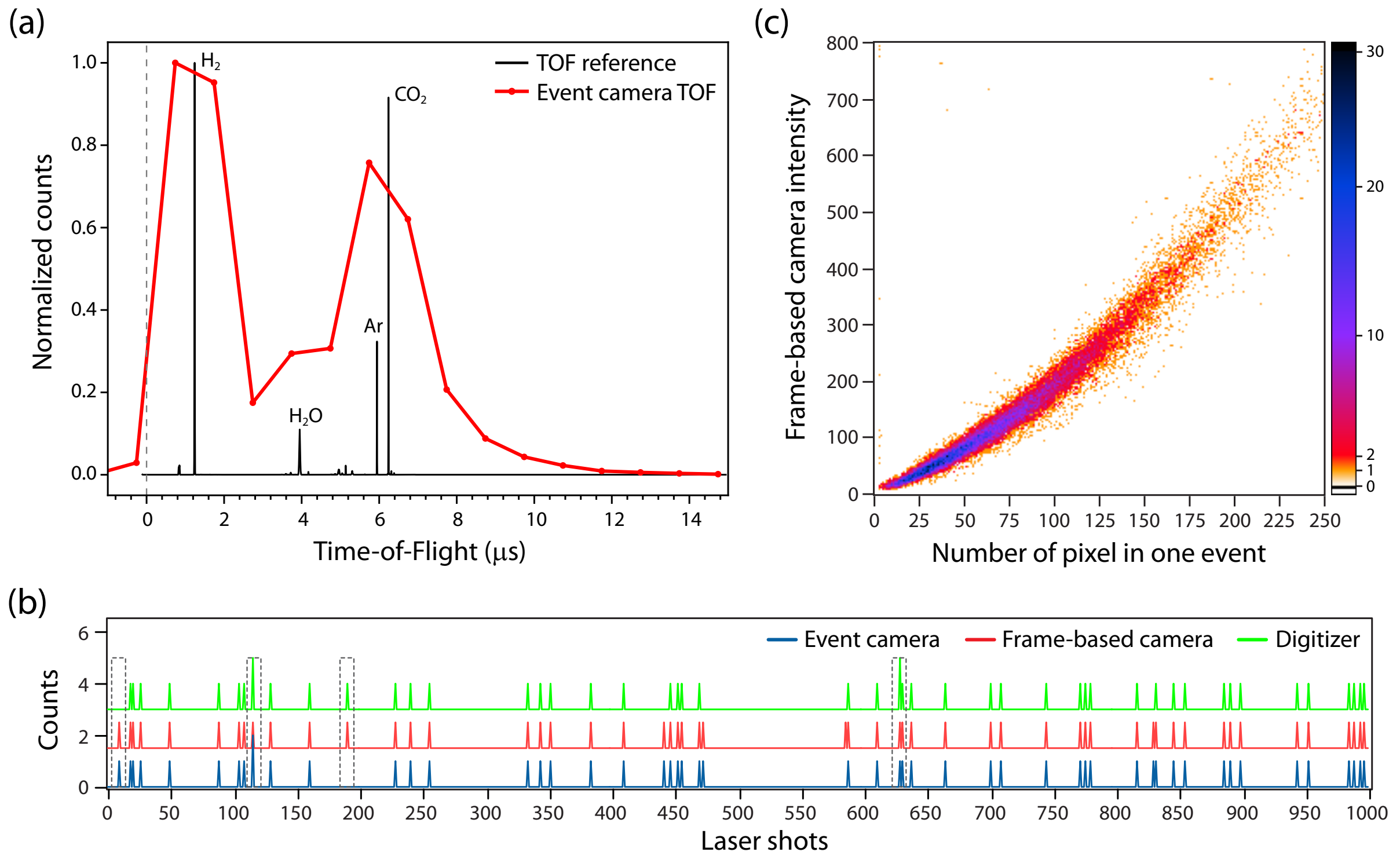

Figure 2: Event camera performance in a 3D VMI system. (a) TOF mass spectra on multiple species. The red curve was obtained from the event camera latency measurement while the black curve was obtained using a digitizer and MCP pick-off. (b) Comparison of event detection efficiency among event camera, frame camera and MCP pick-off+digitizer, showing a comparable efficiency among the three. (c) A strong correlation between the number of pixels activated in each event detected by an event camera and the brightness of the same event measured by a frame camera. The dash boxes indicate the laser shots that have mismatched event detection among the three devices.

The lack of such a capability in event camera would lead to a loss of its multi-hit power for 3D VMI. In TPX3CAM, the intensity was estimated using the time-above-threshold (ToT) value and this does not exist in machine vision event camera. Here in this study, we were able to identify a proxy for the pixel brightness: the number of pixels contained in each ion/electron events. This arises from the fact that a brighter flash always correlates with a larger spatial spread of the flash. This has been confirmed with the two-camera setup. We plot the number of activated pixels in each event against the brightness measured by the frame camera. Fig. 2b shows the strong correlation between the two values, validating the approach.

Fig. 3 shows the 3D momentum measurement performance of the 3D VMI setup, with both the frame camera and the event camera. This was run at 1 kHz and electrons arising from photoemission from perovskites nanoplatelet were detected[21]. Fig. 3a shows the XY images from frame camera with 4× upscaling on the resolution by centroiding[28] while Fig. 3b is the spatial image obtained with the native resolution of the event camera. The event camera shows a better spatial resolution thanks to its much higher pixel count. Fig. 3c shows the Xt distribution in which the TOF (t) measurement was provided by the digitizer and MCP pick-off. The overall performance with the event camera is excellent.

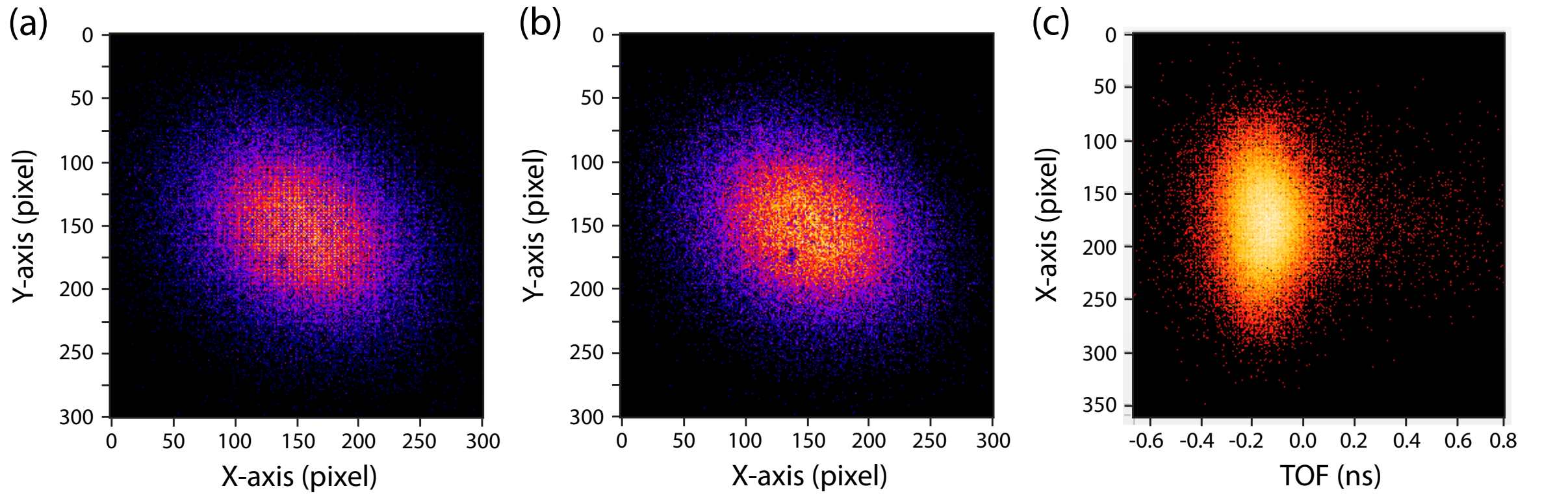

Figure 3: Photoelectron 3D momentum measurement performance. (a) XY spatial distribution obtained using a frame camera. (b) XY spatial distribution obtained using an event camera. (c) Xt image.

Finally, we demonstrate that the system can work with a laser running at 100 kHz repetition rate. Fig. 4a show the latency distribution of electron events. Because the MCP imaging detector employed in the surface photoemission apparatus has aged significantly, its achievable brightness has reduced. This has led to a higher latency. Even in this case, the latency is still below 10 μs, which enabled a successful synchronization between the camera and digitizer events. The 3D momentum distributions XY and Xt are shown in Fig. 4b and c. Interestingly, the delayed emission that was observed at lower repetition rates has largely disappeared, likely due to a higher thermal load on the sample. This is a good example that a new capability of an apparatus will allow observation of new phenomenon. The detailed dynamics is a subject of future investigation.

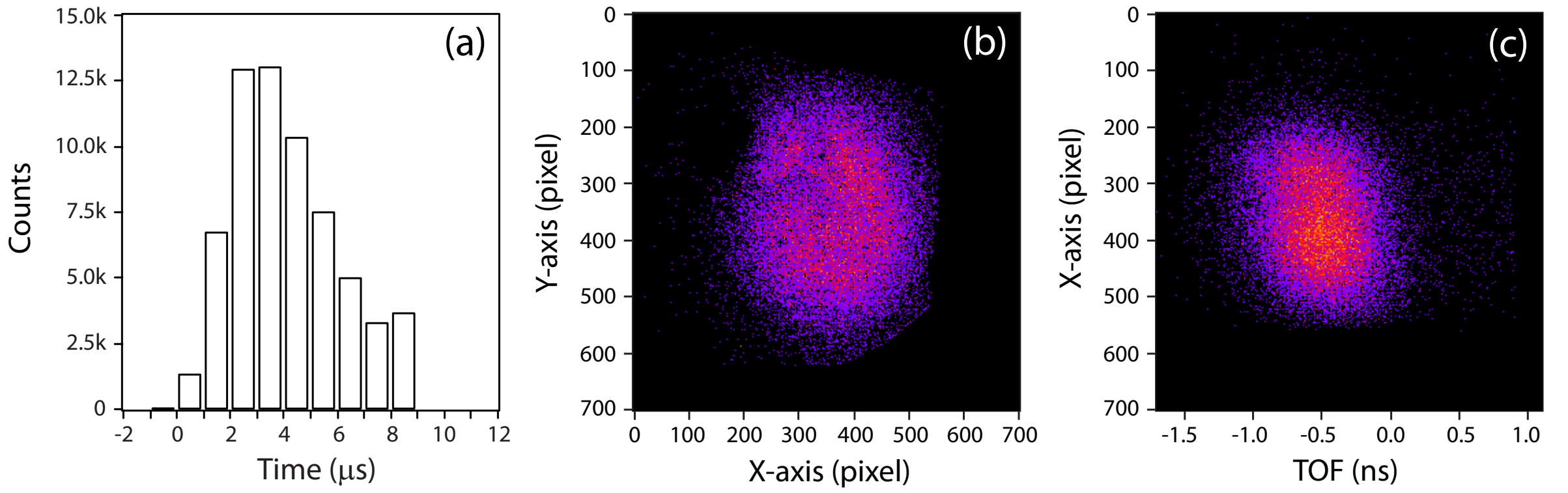

Figure 4: 3D VMI at 100 kHz. (a) A distribution of event latency. (b) XY distribution. (c) Xt distribution.

## Conclusion

To conclude, we have demonstrated an improved 3D VMI setup which employs a machine vision event camera. The unprecedented low latency allows the 3D VMI system to run at 100 kHz, two orders of magnitude faster than what have been demonstrated using frame cameras while providing additional boost on spatial resolution. A higher repetition rate is particularly important for coincidence experiments, where the number of particles detected per laser shot must be kept sufficiently low to minimize false coincidences. Combining high-repetition-rate lasers such as a fiber laser with event camera can be especially useful for investigating weak reaction channels and competing pathways in strong-field dynamics and attosecond spectroscopy, where the yields of some products can be much smaller than those of dominant channels. This improved system will be a powerful tool in the study of chemical dynamics, AMO and condense matter physics, where detecting the 3D momentum of particles is needed.

## Acknowledgement:

Research supported by the U.S. Department of Energy (DOE), Office of Science, Basic Energy Sciences (BES), under Award # DE-SC0020994 (strong field ionization study) and the National Science Foundation, CMI program, under Award No. CHE-2107860 (photoemission study).